\documentclass[graybox]{svmult}

\usepackage{mathptmx}       
\usepackage{helvet}         
\usepackage{courier}        
\usepackage{type1cm}        
\usepackage{mathtools}
\usepackage{cellspace}
\usepackage{multirow}
\usepackage{makecell}
\usepackage{makeidx}         
\usepackage{graphicx}  
\usepackage{color}
\graphicspath{{Figures/}}
\usepackage{multicol}        
\usepackage[bottom]{footmisc}

\usepackage{cite}

\newif\ifcomments
\commentstrue
\definecolor{dkgreen}{rgb}{0,0.6,0}
\definecolor{gray}{rgb}{0.5,0.5,0.5}
\definecolor{mauve}{rgb}{0.58,0,0.82}
\ifcomments
  \newcommand{\noteEL}[1]{\textcolor{dkgreen}{{\scriptsize{EL:}}#1}}
  \newcommand{\pjm}[1]{\textcolor{mauve}{{\scriptsize{pjm:}}#1}}
  
\else
  \newcommand{\noteEL}[1]{}
  \newcommand{\pjm}[1]{}
\fi

\newcommand{\+}[1]{\ensuremath{\mathbf{#1}}}
\makeindex             

\begin{document}

\title*{Exploring concurrency and reachability in the presence of high temporal resolution}
\author{Eun Lee \and James Moody \and Peter J. Mucha}
\institute{Eun Lee \at University of North Carolina, Chapel Hill, \email{eunfeel@email.unc.edu}
\and James Moody \at Duke University, \email{jmoody77@soc.duke.edu}
\and Peter J. Mucha \at University of North Carolina, Chapel Hill, \email{mucha@unc.edu}
}
%
%
\maketitle

\abstract{Network properties govern the rate and extent of spreading processes on networks, from simple contagions to complex cascades. Recent advances have extended the study of spreading processes from static networks to temporal networks, where nodes and links appear and disappear. We review previous studies on the effects of temporal connectivity for understanding the spreading rate and outbreak size of model infection processes. We focus on the effects of “accessibility”, whether there is a temporally consistent path from one node to another, and “reachability”, the density of the corresponding “accessibility graph” representation of the temporal network. We study reachability in terms of the overall level of temporal concurrency between edges, quantifying the  overlap of edges in time. We explore the role of temporal resolution of contacts by calculating reachability with the full temporal information as well as with a simplified interval representation approximation that demands less computation. We demonstrate the extent to which the computed reachability changes due to this simplified interval representation.}


\section{Introduction}
\label{sec:introduction}
Variation in epidemic spreading stems in part from the diversity of temporal contact patterns between subjects,
whether such changes are a direct result of individual state changes (as in, e.g., \cite{DALEY1964,May1987,May1988}) or more general temporal variation (see, e.g., \cite{HOLME2012Physreport, Naoki2016Tempguide}).
For example, the distribution of the lifespan of edges can significantly affect the speed and eventual spread of an infection~\cite{Naoki2013PRL,Holme2015, Li2018}.
The increased availability of detailed, digitized temporal contact patterns supports and accelerates new investigations about the effects of temporal details, including analysis of properties such as fat-tailed inter-event time distributions~\cite{Vazquez2007PRL,Karsai2011PRE,Rocha2011,Gernat2018}. 
Indeed, the `burstiness' of inter-event times can either slow down~\cite{Karsai2011PRE} or speed up dynamics~\cite{Rocha2011,Gernat2018}. 
Meanwhile, such apparently contradictory effects provide a clue that there may be other elements controlling the dynamics beyond the bursty inter-event times. 

Holme and Liljeros~\cite{Holme2015} investigated the changes to the observed outbreak sizes from various selected shifts to the contact histories: ``beginning intervals neutralized" (BIN) shifts all contact pairs to first appear at the same time, ``end intervals neutralized" (EIN) shifts the last contact between all pairs to the same time, and ``interevent intervals neutralized" (IIN) replaces the heterogeneous intervals of contact between a pair to a uniform step size in time (keeping start and end times the same). 
For 12 empirical temporal networks, they found that BIN and EIN resulted in more significant differences in the outbreak size compared with differences obtained from IIN. 
A possible explanation for the relatively larger effect of these BIN and EIN modifications could be in the resulting changes in the concurrency of contacts. That is, by shifting all contacts to start (BIN) or end (EIN) at the same time, there is presumably greater temporal overlap between different contact intervals, augmenting the temporally consistent paths in the network over which the infection may spread.

Further supporting this possible interpretation, Li \textit{et al.}~\cite{Li2018} analyzed the transient behavior of reference models with randomly permuted times that either preserve the lifetimes of edges or of the nodes, focusing on changes in spreading speed according to the selected reference model. Their results demonstrate the dependence between the ability of an edge to help spread the infection and the time interval of its lifetime. Together, these results highlight the importance of the overall time interval over which a given pair is in contact, as opposed to the detailed timings of the contacts in those intervals.

Such studies point to the crucial function of the concurrency of edges in infection dynamics. 
In this chapter, we summarize previous studies related to issues of concurrency and the overall reachability constrained by the network timing details. We then explore the impact of concurrency on reachability by rescaling the start times in a set of empirical temporal networks. 
Using these empirical networks we then demonstrate the accuracy with which reachability is correctly calculated using a simplified interval representation for each edge that ignores the detailed timings of interevent contacts.

\section{Previous studies on concurrency and reachability}
\label{subsec:concurrency}
Although there exist various definitions for concurrency, its essence is clear: the extent of temporal overlap among the contacts. 
The significance of concurrency in a temporal network is immediately obvious for governing the reachable extent of any information or infection. Consider for example a simple situation with only three actors $\{A,B,C\}$ with $B$ and $C$ connected by an edge at some early time and $A$ and $B$ connected at a later time. If the temporal extent of these two edges do not overlap, then there is no way for any infection or information that spreads from $A$ to $B$ to continue on to $C$. However, if the two edges temporally overlap, then $C$ is indeed ``accessible" or ``reachable" from $A$.
The reachable extent allowed by the edge timings in a temporal network immediately impacts the real spread and modeling of an infectious disease, independent of the details of the dynamical process (e.g., SI, SIR, SEIR, complex contagion, etc.). 
The expected size of the maximally reachable set can be quantified by  ``reachability", defined as the fraction of ordered node pairs with  at least one temporally consistent path from the source node to the target node.  Such ordered node pairs are ``accessable."

Because the reachability is an underlying property of a temporal network, independent of the spreading process taking place on that network, and since it naturally constrains all spreading processes on the temporal network, reachability has been used in multiple previous studies~\cite{Moody2002a, PRE2005Holme, PRL2013Hartmut, Moody2016, Armbruster2017, Naoki2017}. 
For example, Holme~\cite{PRE2005Holme} numerically investigated two types of reachability called \textit{reachability time} and \textit{reachability ratio}, to categorize the effectiveness of the real-world contact networks in terms of time and spreading size. 
Lentz \textit{et al.}~\cite{PRL2013Hartmut} also explored accessibility in empirical networks, proposing the use of \textit{causal fidelity}, defined as the fraction of network paths that can be followed by a sequence of events of strictly increasing times. That is, if all of the temporal contact information is agglomerated into a static network (collecting all edges that ever exist in the data but ignoring their timings), causal fidelity is the fraction of paths in this agglomerated static network that are also available in the full temporal network, thereby quantifying how well the static network representation might approximate the full dynamics.

In a related line of inquiry, the effect of concurrent partnerships has been of key interest for  the spread of sexually transmitted diseases (STDs) such as HIV/AIDS~\cite{Morris1995,Kretzschmar1996, Moody2002a}.
Moody~\cite{Moody2002a} emphasized the substantial effect of concurrency on the reachability in an adolescent romantic network, in that reachability plays the role of an upper bound on the expected outbreak size of an infection spreading on the network.
An array of studies have assessed the effect of concurrent relationships for modeling infectious spreading on synthetic networks~\cite{Eames2004,Doherty2006,Miller2017,Naoki2017}. The merit in studying synthetic networks is that it enables researchers to control the network's structural properties and the extent of the concurrent partnerships, which are obviously impossible to control in real-life networks.  
Despite considerable emphasis by different investigators about the role of temporally overlapping contacts, we still lack a general definition of concurrency in that slightly different definitions have been used across these studies.
For example, Gursky and Hoffman~\cite{GURSKI2016Mathbio} assumed concurrency based on the lifetime of an edge, following the definition in~\cite{WATTS1992Mathbio}.
Doherty \textit{et al.}~\cite{Doherty2006} defined concurrency as the proportion of subjects engaging in concurrent relationships within a population.
Onaga \textit{et al.}~\cite{Naoki2017} set concurrency as a fixed number of connections of an individual in time. 

Onaga \textit{et al.}~\cite{Naoki2017} proposed a theoretical framework for the epidemic threshold induced by concurrency. 
In general, a low epidemic threshold can indirectly indicate a high probability of infection prevalence. Further, the relationship between concurrency and the epidemic threshold can help explain the relationship between concurrency and reachability. 
Onaga \textit{et al.} defined the concurrency as a fixed number of links emanating from a node in unit time, and the activation of the links are decided by a node's activity level. 
The activity level is drawn from uniform and power-law  probability distributions. 
Then, they applied the analytically tractable activity-driven model. 
Given the star-like network in unit time, the authors derived differential equations for a SIS model to estimate the epidemic threshold. 
They compared the analytically derived threshold to the numerically estimated threshold, confirming a close match. 
From the results, the authors found that the transition of the epidemic threshold depended strongly on the extent of the concurrent connections. 
The results, again, stress the importance of concurrency in adjusting infectious potential. 

In the present work, we are motivated by the framework investigated by Moody and Benton~\cite{Moody2016}, which focused on the roles of concurrency and structural cohesion.
They performed numerical experiments simulating edge start times and durations on network structures sampled with a four-step random walk from a collaboration network. Moody and Benton thus obtained 100 sample networks with which they explored different levels of structural cohesion, defined as the average number of node-independent paths between node pairs~\cite{Moody2003}. 
The authors controlled the concurrency --- quantified by the fraction of connected edges whose temporal intervals overlap in time --- by adjusting the distributions of the start times and durations of the edges. 
Given the sample networks with random start and duration times on each edge, they then measured reachability as a function of concurrency and modeled the relationships with general linear regression models. 
Moody and Benton showed that the concurrency and structural cohesion both affect reachability in their examples, finding that the role of concurrency is particularly important in low structural cohesion networks because a slight increase in concurrency sharply increases the number of accessible node pairs (that is, ordered node pairs connected via temporally consistent paths). 
When one considers that networks of low structural cohesion are common in many sexual contact networks, Moody and Benton's findings stress the importance of concurrency for STD transmission.

Recently, Lee \textit{et al.}~\cite{LEE2019} developed a tree-like model approximations for the relationship between concurrency and reachability, to further elucidate numerical results like those of Moody and Benton~\cite{Moody2016}.
Lee \textit{et al.}\ compared their approximations with numerically-computed reachability in temporal networks obtained with simulated edge timings on various network structures: balanced and unbalanced trees, Erd\H{o}s-R\'enyi (ER) networks, exponential degree distribution networks, and four of the sampled networks highlighted in~\cite{Moody2016}.
Because of the nature of their tree-like assumptions, these models well approximate the relationship between reachability and concurrency for small values of structural cohesion, doing particularly well also at small values of concurrency. But their existing models do not do as well in the presence of larger numbers of available alternate paths between nodes.
Nevertheless, this study further demonstrates how the overall level of reachability emerges through an interplay between concurrency and structural cohesion.

\section{Effects of concurrency: empirical examples}
In the remainder of this chapter, we focus on a set of empirical examples to further explore reachability and concurrency, complementing the results in~\cite{Moody2016} and \cite{LEE2019}. 
As part of our exploration, we transform the detailed contact time information of the edges into an interval representation wherein the distinct contacts between each connected node pair is instead represented simply by a start time (the first observed contact) and an end time (the last contact).
That is, we treat each edge as if it was present for the entirety of the interval between the first and last observed contacts.
We then measure concurrency and reachability on this simpler interval representation. To perform numerical experiments under different values of concurrency, we modify these time intervals by rescaling the total range of the start times in the temporal network while keeping the duration of each edge constant.
By doing so, we investigate how concurrency affects reachability and examine whether reachability on the simpler interval representation matches that measured from the original contact times.

The basic characteristics of the four example empirical networks used in this study are described below and in Table~\ref{tb:dataset}. In the following subsections, we then describe the transformation to the interval representation, the measurement of concurrency and our method for modifying it in our present simulations, and the calculation of reachability. Using the empirical examples, we then demonstrate the impact of concurrency on reachability as well as the relative accuracy of computing reachability with the simplified interval representation in these examples.

\subsection{Data}
We used four empirical networks in the present study. The first data set (denoted ``High School" here) contains the temporal network of contacts between students in a high school in Marseilles, France, including contacts between the students in five classes during a seven day period (from a Monday to the Tuesday of the following week) in November 2012~\cite{Highschool2014Plosone}.
The network includes $N=180$ nodes and $M_c = 45,047$ distinct contacts between $M_d = 2,220$ different node pairs (that is, yielding $M_d$ different edges in the interval representation).
The time resolution of the measured contacts is $\Delta=20$ seconds.

The second data set (``Conference") corresponds to the contacts among attendees of the ACM Hypertext 2009 conference~\cite{Conf2011Theobio}. The conference spanned 2.5 days, with the network sampled every $\Delta=20$ seconds. The network consists of $N=113$ attendees and $M_c = 20,818$ contacts between $M_d = 2,196$ node pairs.

The third data set (``DNC Email") is the Democratic National Committee email network, as hacked from the DNC in 2016 (data available online at {http://konect.uni-koblenz.de/networks/dnc-temporalGraph}). Nodes in the network correspond to persons, with each contact along an edge representing an email sent to another person. Although the data are originally directed, we treat edges here as undirected for simplicity. The network includes $N=1,891$ nodes and $M_c = 39,264$ email contacts, connecting $M_d = 4,465$ node pairs. 

The fourth data set (``Brazil") is a sexual contact data set obtained from a Brazilian web forum  exchanging information about sex sellers between anonymous, heterosexual, male sex buyers between September 2002 and October 2008~\cite{Rocha2010pnas}. 
In this web forum, male members grade and categorize their sexual encounters with female escorts in posts using anonymous nicknames.
From the posts, Rocha \textit{et al.}~\cite{Rocha2010pnas} constructed a network connecting every community member (sex buyer) to an escort. The time information of the posts are used here as the temporal contact between a seller and buyer. The entire network's size is $16,730$. However, to save computational cost, we ignored temporal contacts that occurred during the first $1,000$ days of the data. Additionally, whereas the original data is resolved at the level of days, we down-sampled the resolution of the contacts to $\Delta = 5$ days. As a result, the data we consider includes $N=6,576$ nodes and $M_c = 8,497$ distinct contacts along $M_d = 8,056$ edges.

\begin{table}[ht]
\begin{center}
\begin{tabular}{ |p{3.0cm}|p{1.7cm}|p{1.7cm}|p{1.7cm}|p{1.7cm}| }
 \hline
 \makecell[c]{\textbf{Name of the data}}
 & \makecell[c]{\textbf{$N$}} &\makecell[c]{\textbf{$M_c$}}
 & \makecell[c]{\textbf{$M_d$}} & \makecell[c]{\textbf{$\Delta$}}\\
 \hline
 \makecell[c]{High School}  & \makecell[c]{180} & \makecell[c]{45,047} & \makecell[c]{2,220} & \makecell[c]{20\,s}\\ 
 \makecell[c]{Conference} & \makecell[c]{113} & \makecell[c]{20,818} &\makecell[c]{2,196} & \makecell[c]{20\,s}\\ 
 \makecell[c]{DNC Email} & \makecell[c]{1,890} & \makecell[c]{39,263} & \makecell[c]{4,463} & \makecell[c]{1\,s} \\ 
 \makecell[c]{Brazil} & \makecell[c]{6,576} & \makecell[c]{8,497} & \makecell[c]{8,056} & \makecell[c]{5\,days}\\ 
 \hline

\end{tabular}
\end{center}
\caption{Four empirical temporal networks used in the present work. The networks are of size $N$ (number of nodes) with $M_c$ distinct temporal contacts between $M_d$ different node pairs (the number of edges). The resolution of the temporal contacts is denoted by $\Delta$.}\label{tb:dataset}
\end{table}

\subsection{Change to the interval representation}
\label{subsec:intervalrepresentation}
\begin{figure}[!htp]
\centering
\includegraphics[width=0.7\linewidth]{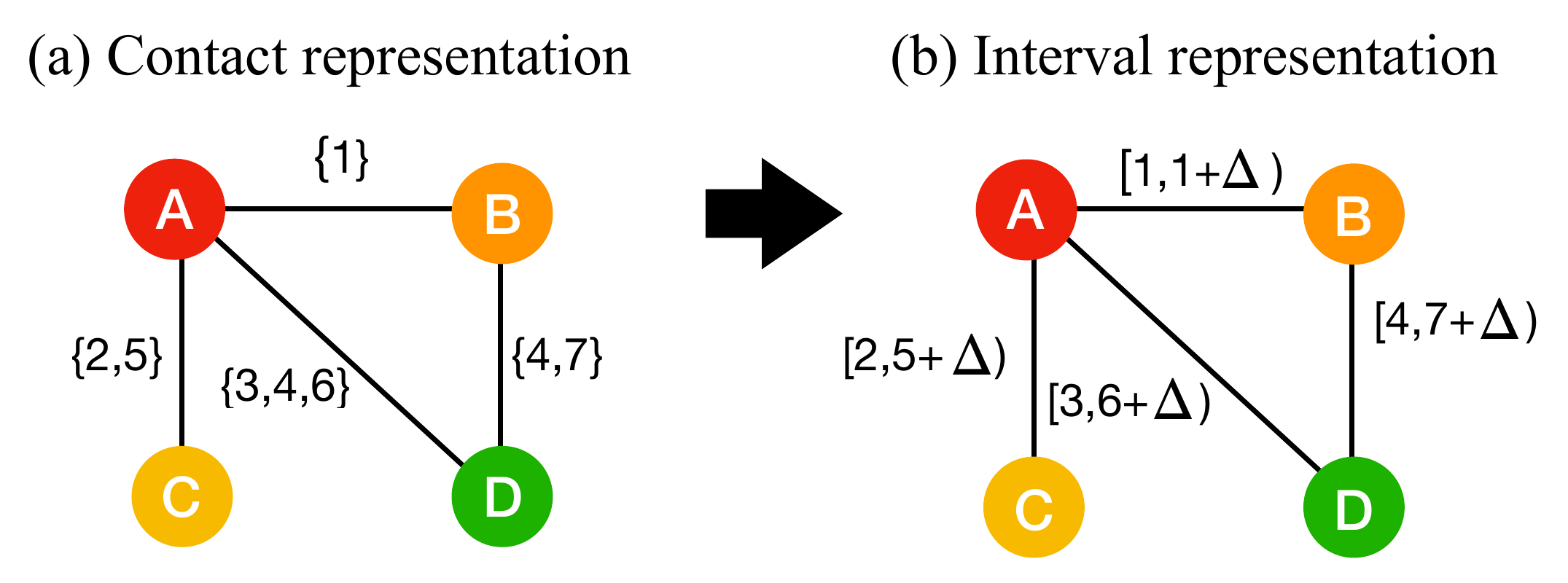}
\caption{A toy temporal-network example represented by (a) distinct contacts and (b) the corresponding interval representation. The interval of each edge starts with the first observed contact. To account for temporal resolution, we set a strict inequality (open interval) end time equal to the last observed contact plus the temporal resolution $\Delta$.}
\label{fig:change_interval}
\end{figure}

The empirical networks include detailed temporal contact patterns like that represented in Fig.~\ref{fig:change_interval}(a): an edge representing the connection between nodes $i$ and $j$ has potentially several time stamps that represent the distinct contacts between $i$ and $j$. The detailed transmission of any infection occurs during these contacts. However, instantaneous contacts are not necessarily the best way to think about concurrency in these relationships. Consider the motivation to study the spread of STDs: the appropriate notion of concurrency isn't that the contacts occur at precisely the same time, only that they are interleaved in time. 

As such, in our present investigation of concurrency and reachability we employ a simplification obtained by transforming the temporal details in the contacts into an interval representation, keeping only the start- ($t_s$) and end time ($t_e$) of each edge, as shown in Fig.~\ref{fig:change_interval}(b). In panel (b), the contacts of the edge between A and D --- which includes contacts at times \{3,4,6\} (see. Fig.~\ref{fig:change_interval}(a)) --- are converted to the time interval $[3,6+\Delta)$. In this transformation, we explicitly add the time resolution $\Delta$ to the last contact time so that every edge includes a non-zero time interval even if it represents only a single contact at time $t$ [for example, the edge (A,B) in panel (a)]. Consistent with this addition, in our convention the edge only persists for times strictly less than the end time of the interval (open interval on the right).  

\subsection{Measuring and controlling concurrency}
In the present work, we measure concurrency as the fraction of edge pairs that overlap in time. In so doing, we first emphasize that the key mechanism through which concurrency plays out is at the level of connected edges (that is, two edges that share a common node). However, in simulations where edge timings are independent and identically distributed, such as those in \cite{Moody2016} and \cite{LEE2019}, the expected measurement of concurrency over all edge pairs is equivalent to that over the subset of connected edge pairs. In practice, in the real world, whether one more naturally defines concurrency over all edge pairs or only connected edge pairs may be directly determined by the nature of surveyed information. For example, if distributions of start times and durations of edges are measured, then the resulting estimate is effectively over all pairs. In contrast, if participants are directly queried about their numbers of concurrent relationships, then the restriction to connected edge pairs may be more natural. For the purposes of the present article, we measure concurrency as the fraction of all edge pairs that overlap in time. 
This definition enables us to more easily analyze the effect of concurrency on the reachability, particularly in developing models for the effect as in \cite{LEE2019}. 

\begin{figure}[!htp]
\centering
\includegraphics[width=0.95\linewidth]{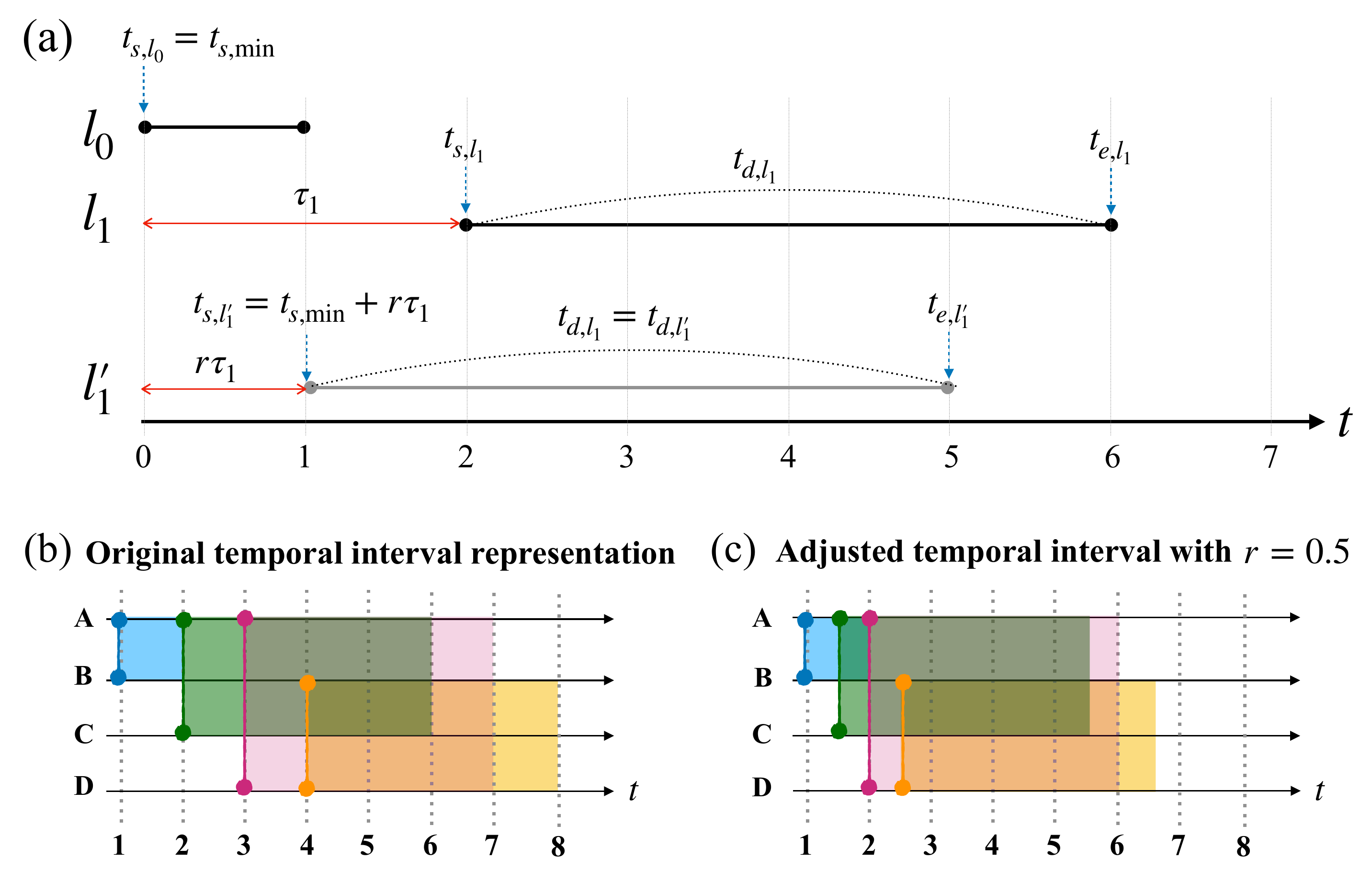}
\caption{Controlling concurrency in our toy example. (a) The very first link to appear, $l_0$, sets the minimum starting time in the empirical data, $t_{s,\rm{min}}=t_{s,l_0}$, and the temporal axis in panel (a) is expressed in time since $t_{s,\rm{min}}$. 
A later link, $l_1$, has an empirical start and end time: here $t_{s,l_1}=2$ and $t_{e,l_1}=6$, that is, with duration $t_{d,l_1}=4$. 
The grey line $l'_1$ represents the new $l_1$ after rescaling the start time distribution by $r=0.5$, with new start time $t_{s,l'_1}$ obtained by pulling the original start time forward by a factor of $r$. 
In this rescaling of the start times, the duration of the edge remains the same.
Under this rule, the original intervals in panel (b) with time resolution $\Delta=1$ change under the rescaling $r=0.5$ to those in panel (c). 
The concurrency of the intervals in the two panels are $C_{r=1.0} = 3/6 = 0.5$ and $C_{r=0.5} = 4/6 = 0.67$. Further increases in $r$ increase $C$ here until $C=1$ for $r<1/3$.}
\label{fig:control_conc}
\end{figure}

In our numerical experiments, we control the concurrency by rescaling the edge start times without changing their durations. 
We identify the minimum start time $t_{s,l_i}$ of each edge --- that is, each pair of nodes that are ever in contact --- where $l_i$ indicates the $i$th edge, $i\in [0,1,\dots,L-1]$, and $L$ is the total number of edges. (Connecting the notation of this section to our data analysis, we note that $L=M_d$.) 
For notational convenience, we identify the very first start time $t_{s,\rm{min}} = \min t_{s,l_i}$ among all edges and define $\tau_i = t_{s,l_i} - t_{s,\rm{min}}$. 
We can then rescale the distribution of these start times with the parameter $r$ by $t_{s,l'_i} = t_{s,\rm{min}} + r\tau_i$, as depicted in Fig.~\ref{fig:control_conc}(a). 
Meanwhile, we maintain the duration of each edge with $t_{e,l'_i} = t_{s_l'i}+t_{d,i}$ where $t_{d,i}=t_{e,i}-t_{s,i}$ is the edge duration.
For example, when we set $r=0.5$, the interval representation of the original timings depicted in Fig.~\ref{fig:change_interval}(b) and Fig.~\ref{fig:control_conc}(b) (corresponding to $r=1$) shift to the intervals in Fig.~\ref{fig:control_conc}(c). 
In particular, note that the edges in panel (c) overlap each other more than the original edges in panel (b).

\subsection{Measuring reachability}
\begin{figure}[!htp]
\centering
\includegraphics[width=0.7\linewidth]{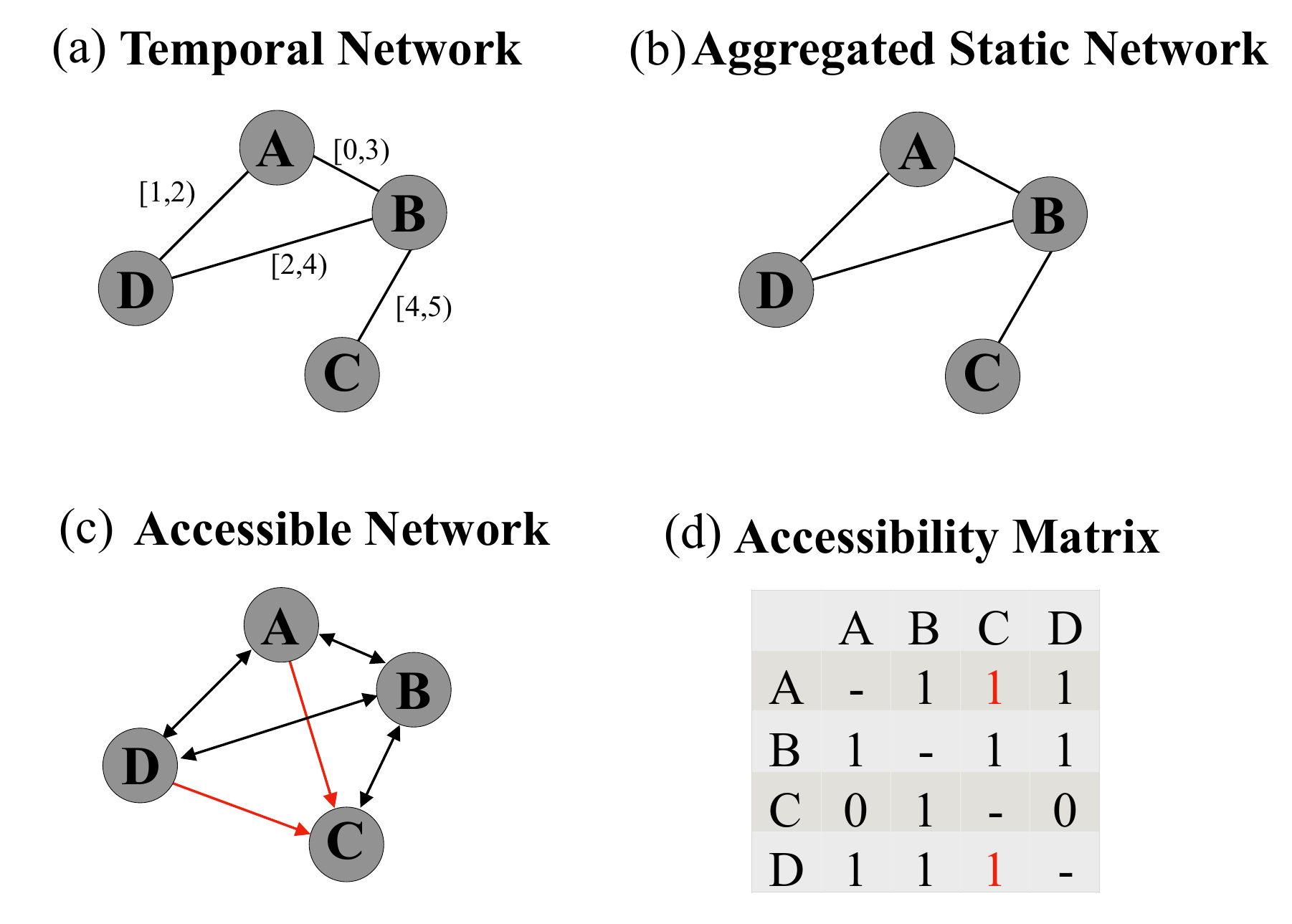}
\caption{Schematic representations for establishing accessibility. (a) Each edge in the network of nodes $\{A,B,C,D\}$ is denoted by a start and end time, e.g., the contact between $A$ and $D$ starts at $t=1$ and continues until (just before) $t=2$. (b) The static network representation aggregates all contacts that ever appear in the temporal network. (c) The corresponding directed graph of accessibility demonstrates that asymmetric accessibilities (red arrows) are possible.
(d) The accessibility matrix encodes whether node $j$ is accessible from node $i$.
}
\label{fig:schem}
\end{figure}

Given the interval representation of each edge, we can evaluate the average reachability of the network corresponding to that representation. 
Consider the (different) toy example in Fig.~\ref{fig:schem}(a).
In general, all direct contacts such as those connecting (A,B) yield accessible node pairs.
Additional pairs like (A,C) and (D,C) in Fig.~\ref{fig:schem}(c) are accessible because of the temporal ordering of the edges.
For example, an infection starting from D at $t=1$ can reach C by either moving first to A and then on to B, or moving directly to B, and then having B infect C at a later time.
However, an infection seeded at C cannot ever reach A or D because of an absence of available connections after the appearance of the (B,C) link at $t=4$. That is, neither A nor D is accessible from C. 
(We again emphasize our convention of open intervals on the right, so in the example here the edge between B and D disappears immediately before the edge between B and C starts.)
We identify the accessible ordered node pairs with the elements of the accessibility matrix $\+R$, with $R(i,j)=1$ if and only if $j$ is accessible from $i$, as shown in Fig.~\ref{fig:schem}(d). 

To quantify the overall average accessibility across the whole network, we calculate the reachability $R$ as the density of the accessibility graph (i.e., the density of the off-diagonal elements of the accessibility matrix),
\begin{equation}
\label{eq:R}
    R = \dfrac{1}{N(N-1)}\sum_{\substack{i\neq j}} R(i,j)\,.
\end{equation}

To calculate the accessibility matrix $\+R$ and reachability, we follow the same steps as in \cite{LEE2019}, generating temporal layers corresponding to the moment immediately before the end of each edge:
\begin{enumerate}
\item Sort edges by their end times $t_{e,l_w}$. Here, $w \in [0,L-1]$ indexes edges by their end time in ascending order and $L$ is the total number of edges. (We again note $L=M_d$ here.)
For example, $l_0$ is the edge with the earliest end time and $l_{L-1}$ is the last edge to end. (Breaking ties is unimportant for calculating reachability, except it can be used to reduce the number of calculations here, under an appropriate change of notation.)
\item Construct the adjacency matrix $\+T_w$ for the $w$th temporal layer by including edge $l_w$ and all other edges $l_{w'}$ with $w'>w$ (that is, that end after the $w$th edge) that are also present just before the end time of the $w$th edge. That is, $\+T_w$ includes $l_w$ and all $l_{w'}$ satisfying both $t_{s,l_{w'}} < t_{e,l_w}$ and $t_{e,l_{w'}}\geq t_{e,l_w}$.
\item By repeating step 2, the full set of temporal layer matrices $\+{T}_0,\+T_1, \dots, \+T_{L-1}$ may be prepared.
\item Multiply the matrix exponentials of each temporal matrix:
${\+R} = \prod_{w=0}^{L-1} e^{\+T_w}$.
\item Binarize ${\+R}$: For all $R_{ij} >0$, set $R_{ij}=1$. 
\item By using Eq.~\ref{eq:R}, evaluate the average reachability ${R}$.
\end{enumerate}

The matrix exponentials in step 4 provide a simply-expressed formula to indicate connected components within each temporal layer. Multiplying the matrix exponentials from consecutive layers yields (after binarizing) the reachable network associated with that set of layers. While the matrix exponential works conveniently for small data sets, for larger networks a more computationally tractable procedure is to instead directly calculate the connected components of $\+T_w$ and replace the matrix exponential in step 4 with the binary indicator matrix whose elements specify whether the corresponding pair of nodes are together in the same component at that time. In practice, steps 3 and 4 can be trivially combined to separately consider each temporal layer in isolation from the others. That is, with this procedure the calculation can be performed without forming and holding the entire multilayer representation at one time (cf.\ breadth-first search on the full multilayer network). For even larger networks whose adjacency matrices must be represented as sparse matrices in order to even fit in memory, the corresponding accessibility graph could instead be constructed one row at a time, updating the running average of the density $R$ to calculate the overall reachability.

The above-described procedure for calculating reachability for the interval representation can be used for the full temporal contact information with only minor modification.
Instead of sorting edges by their end times, the reachability due to detailed temporal contacts proceeds by taking each possible contact interval as a separate temporal layer. The adjacency matrix and its exponential (or component indicator) is computed for each temporal layer, multiplying the later together as in step 4 except that the index runs over all unique contact times.

\subsection{Reachability with concurrency}
\label{subsec:reach_empinets}
\begin{figure}[!htp]
\centering
\includegraphics[width=0.8\linewidth]{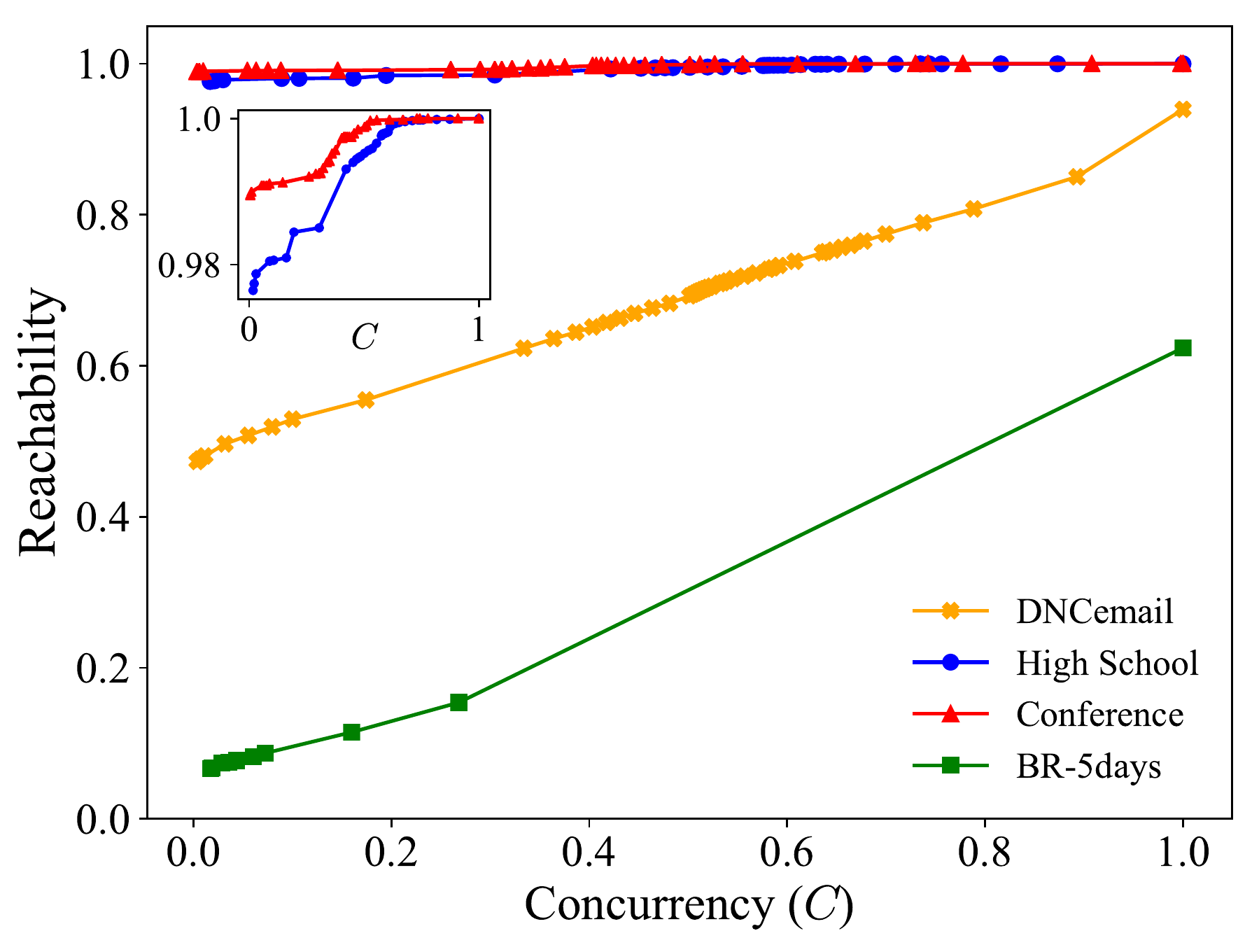}
\caption{Reachability of the empirical networks as a function of concurrency. Different levels of concurrency have been obtained here by rescaling the distribution of start times in the original data sets. The inset zooms in on the small deviations of reachability from $1$ for the High School and Conference examples. Note that the lines here are only to connect the data points; the lines do not represent a functional relationship.}
\label{fig:conc_reach_empi}
\end{figure}

After transforming the temporal contact information into the interval representation (as described in Sec.~\ref{subsec:intervalrepresentation}), we measure reachability versus concurrency in the empirical network data sets. We use the parameter $r$ to rescale the start times in order to vary concurrency. In general, reachability increases with increasing  concurrency (see Fig.~\ref{fig:conc_reach_empi}) although the details of this relationship between reachability and concurrency depends on the topology of the underlying network. 

The High School and Conference examples display only a slight increase in reachability with increasing concurrency for the simple reason that reachability is already so high at zero concurrency. There are so many alternative paths between node pairs in these two networks that almost all pairs have at least one temporally consistent path, even for very small concurrency, and so reachability is almost always close to $1$ in these cases (see the inset of Fig.~\ref{fig:conc_reach_empi}). The interval representation of the original edge timings --- that is, before we rescale the distribution of start times with the $r$ parameter described above --- corresponds to $C=0.25$ (High School) and $C=0.4$ (Conference). 

In contrast, the DNC email network has larger concurrency in its original edge timings ($C=0.6$) but much smaller reachability, as seen in Fig.~\ref{fig:conc_reach_empi}. 
Even as the start time distribution is compressed (small $r$), to make the concurrency approach $1$, the reachability only approaches $0.94$ (not $1$). 
This apparent discrepancy is because the data set includes separate connected components. 
That is, increasing concurrency all the way to $1$ reduces the question of accessibility to connected components in the temporally-aggregated network, with reachability then equal to the fraction of node pairs in the same connected component.
In the temporally-aggregated DNC email network, the largest connected component is of size $1,833$, with another component of size $58$. 

Similarly, the relatively small value of reachability for the Brazil network as $C \to 1$ is because the largest connected component includes $5,193$ (of the $8,056$ total) nodes. At $r=1$, the Brazil data has concurrency $0.0172$. As such, we can see that increasing the level of concurrency (that is, $r<1$) can dramatically increase the reachability for this network.

We note in particular the behavior of the High School and the Conference data sets in having reachability values near 1 for all values of concurrency. We point the interested reader back to \cite{Moody2016} and \cite{LEE2019}, where the important role of structural cohesion in the temporally-aggregated graph is demonstrated. We note that the structural cohesion calculated \cite{Newman2001} for these two networks are $18.3$ (High School) and $28.5$ (Conference), quantifying the large number of node-independent paths typically available in these networks. In contrast, the structural cohesion of the DNC email network is $1.28$, directly quantifying that it is much more tree-like, and as such there are typically few (or in many cases no) available detours between nodes.
Similarly, the structural cohesion of the Brazil network is $1.21$.
Given the particularly large values of structural cohesion for the High School and Conference networks, reachability values near $1$ are not surprising, even as concurrency approaches zero.

\subsection{Accuracy of reachability from the interval representation}
\begin{figure}[!htp]
\centering
\includegraphics[width=0.98\linewidth]{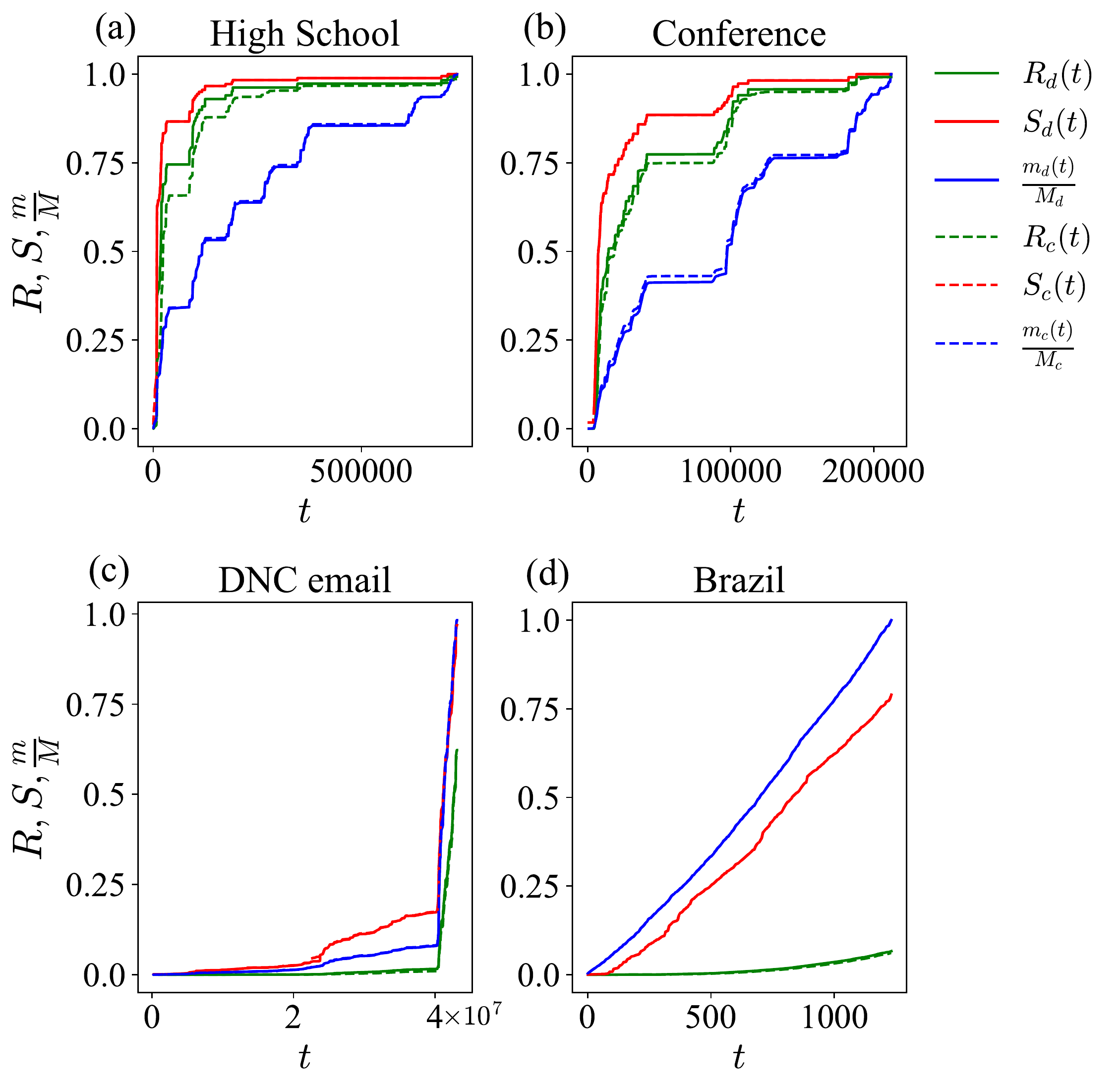}
\caption{Temporal traces of reachability ($R$), the size of the largest component ($S$) and the normalized edge count ($m/M$) in four empirical networks as calculated from the contacts (subscripted with $c$, plotted as dashed lines) and the interval representation (subscripted with $d$, solid lines). In many cases, the dashed lines are not distinguishable from the corresponding solid lines. 
}
\label{fig:empi_time_trace}
\end{figure}

To further explore reachability and its dependence on the temporal details of the contacts, we calculate reachability in the four empirical temporal networks, tracing the change in reachability over time in the original data sets (i.e., without modifying start times).  
Figure~\ref{fig:empi_time_trace} demonstrates the different increasing trends of reachability with time $t$ across these networks, setting $t=0$ in the figure at the appearance of the very first contact. 
Figure~\ref{fig:empi_time_trace} also visualizes this increase in reachability relative to the number of edges $m(t)$ that have appeared by that time (i.e., the number of distinct node pairs that have had contact by that time). 
The figure includes calculations using the original contact times as well as those from the interval representation wherein each edge is assumed to be present for the full duration from its first appearance to its last. 
We use subscripts to distinguish between the calculations using the distinct temporal contacts ($c$) versus the interval representation ($d$, indicating each edge is assumed to be present for its total duration).

\begin{figure}[!htp]
\centering
\includegraphics[width=0.98\linewidth]{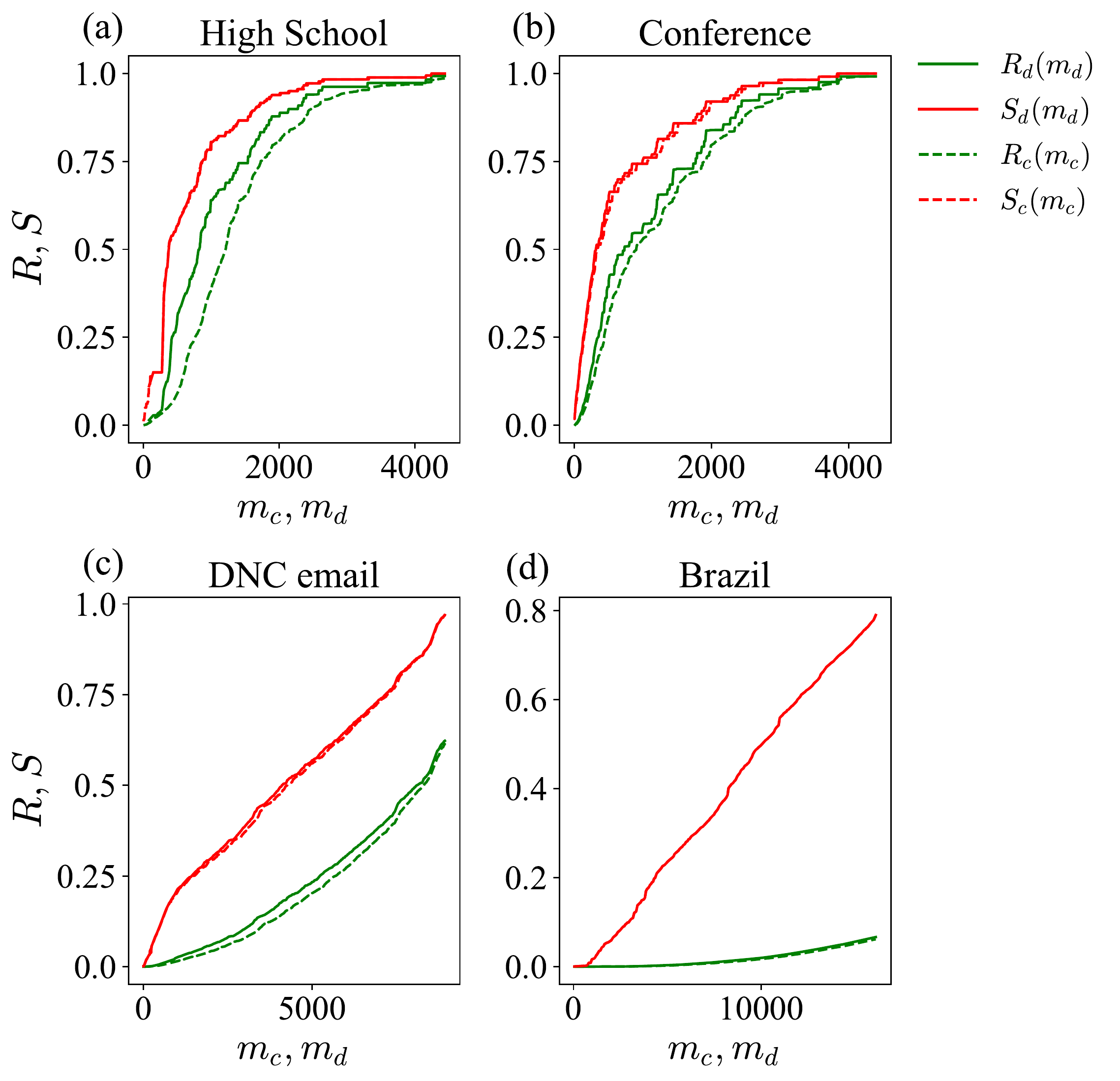}
\caption{Temporal traces of reachability ($R$) and the size of the largest component ($S$) in the four empirical networks, previously plotted in Fig.~\ref{fig:empi_time_trace} are re-plotted here versus the number of edges that have appeared to that point in time. Once again, some of the dashed lines corresponding to calculations with the contacts are indistinguishable from the solid lines obtained from the interval representations.
}
\label{fig:empi_density_trace}
\end{figure}

In addition to reachability ($R_d$, $R_c$), Fig.~\ref{fig:empi_time_trace} includes the largest connected component size ($S_d$, $S_c$) and normalized edge count ($m_d/M_d$, $m_c/M_c$) as a function of time, considering all edges that have appeared since the very first contact. 
For ease of comparing different time scales, we re-plot these results for reachability and the size of the largest connected component  versus the edge densities in Fig.~\ref{fig:empi_density_trace}.
As observed in the figures, the differences between the calculated values based on full contacts versus the interval representation are relatively small, and in many cases barely distinguishable.

Of course, any error in computing the accessibility of an ordered node pair in the interval representation can only overestimate reachability. That is, an ordered node pair identified as accessible in the full contact representation is necessarily also accessible in the interval representation. However, it is possible that particular paths that appear to be temporally consistent in the interval representation do not actually have an allowed set of distinct contacts. 
That said, because our reachability calculation in the interval representation only computes results at the end times of edges, a new edge that appears (the node pair have their first contact) at time $t$ does not get accounted for in the interval representation until the first end time that occurs after $t$. (At that time, this new edge is accounted for, even if its end time is much later.)
By showing the results of both calculations, we demonstrate how accurately the interval representation describes reachability in these examples, with good agreement throughout Figs.~\ref{fig:empi_time_trace} and \ref{fig:empi_density_trace}. 

In line with the very high structural cohesion of the High School and Conference networks, we observe very sharp increases in reachability at early times, with reachability values only slightly behind the fraction of nodes in the largest connected component.
In contrast, we observe in the figure that the reachability of the DNC email and Brazil networks increase more slowly with time, even after redisplaying reachability versus normalized edge count.
Remarkably, reachability calculated from the interval representation deviates only slightly from the full calculation using the complete temporal contact details. The most notable differences between the two calculations apparent in Fig.~\ref{fig:empi_density_trace} is in the High School data, with the interval representation slightly overestimating reachability through its increase over time. A smaller overestimate is also apparent in the panels for the Conference and DNC email networks.

Considering the importance of reachability as the average of the maximum possible outbreak size (averaging over ``patient zero" source nodes), these results provide hope that reachability can be well estimated from the simpler interval representation in most cases, even though the detailed dynamics of a spreading infection surely varies between the true contacts and the interval representation.

\section{Final Remarks}
\label{sec:discussion}
The details of edge timings in a temporal network can affect the speed and extent of the spread of diffusive dynamics such as infections or information propagation on the network. But because including temporal details greatly increases the complexity of the system, there has been a much greater amount of study and successful modeling of spreading processes on static networks. With ever greater emphasis on temporal network data, focusing on the role of concurrency appears to be one productive way to accurately summarize the population-level effects of the edge timing details. 
We here collected references to some previous studies related to the impact of concurrency on spreading processes, including in particular the relationship between concurrency and the average reachability in the temporal network. 
We have further demonstrated this relationship by calculation of reachability on empirical examples, rescaling the start time distributions in the original edge timing data to consider different levels of concurrency and reachability.

In so doing, we also compare the calculation of reachability on the full contact information against that using a simplified interval representation that treats each edge as present for the entire interval between the appearance of its first contact and its last.
We demonstrate with these examples that the level of reachability calculated in the interval representation is nearly identical to that calculated with the full temporal contact information. We note that this result is similar at least in spirit to the findings of \cite{Holme2015} where the detailed inter-event timings did not affect the results in their model simulations as much as the start times and end times.

In terms of the temporal trace of reachability, the High School and Conference networks show simultaneous increase of reachability with the size of the largest connected component at early times. In contrast, the DNC email and Brazil networks display a much slower increase in reachability, lagging behind the connected component size, and the reachability in these networks remains relatively low. 
We confirm with these empirical examples that the effect of concurrency can be quite large in some networks, as seen for the DNC email and Brazil networks. 

The importance of concurrency was first identified in the context of the spread of HIV~\cite{Morris1995}.
Conflicting observational works at the national and individual levels have since raised questions about the value of concurrency in the public health context (see, e.g., \cite{Lurie_Rosenthal_2010,Mah_Halperin_2010,Morris_Epstein_Wawer_2010,Epstein_Morris_2011}), but most of this work misunderstands the necessary relation between reachability and diffusion risk highlighted here (and in, e.g., \cite{Moody2016,LEE2019}).
Whereas increased concurrency increases temporal path accessibility, and this increased reachability must increase diffusion potential, the amount of increase in reachability depends on other network factors, as we have demonstrated.
While we have no data to speak directly to these questions about the value of concurrency in the public health context, our results suggest that one contributing factor might be high variance in the levels of structural cohesion in the underlying networks.
As such, by analyzing the extent of concurrency in a temporal network and its impact on reachability given the structural properties of the underlying network, one might be able to better choose between different intervention strategies to best mitigate the spread of an infectious disease or enhance the extent of positive behaviors.
We hope this chapter serves to gather relevant previous studies and motivate future work.

\begin{acknowledgement}
We thank Petter Holme and Jari Saram{\"a}ki for the invitation to write this chapter. 
Research reported in this publication was supported by the Eunice Kennedy Shriver National Institute of Child Health \& Human Development of the National Institutes of Health under Award Number R01HD075712. 
Additional support was provided by the James S. McDonnell Foundation 21st Century Science Initiative - Complex Systems Scholar Award (grant \#220020315) and by the Army Research Office (MURI award W911NF-18-1-0244). 
The content is solely the responsibility of the authors and does not necessarily represent the official views of any supporting agency.
\end{acknowledgement}
%

\bibliographystyle{spphys}

\end{document}